\documentclass{article}

\usepackage{PRIMEarxiv}

\usepackage[utf8]{inputenc} 
\usepackage[T1]{fontenc}    
\usepackage{hyperref}       
\usepackage{url}            
\usepackage{booktabs}       
\usepackage{amsfonts}       
\usepackage{nicefrac}       
\usepackage{microtype}      
\usepackage{lipsum}
\usepackage{fancyhdr}       
\usepackage{graphicx}       
\graphicspath{{media/}}     
\usepackage{natbib}
\usepackage{graphicx}
\usepackage[figuresright]{rotating}
\usepackage{amsmath}
\usepackage{amsthm}
\usepackage{amssymb}
\usepackage{tablefootnote}
\usepackage{float}
\usepackage{lscape}
\usepackage{caption}
\usepackage{chngcntr}
\usepackage{pdflscape}
\usepackage{appendix}
\usepackage{dsfont}
\usepackage{algorithm}
\usepackage{algpseudocode}
\usepackage{multirow}
\usepackage{url}
\usepackage[compact]{titlesec}
\usepackage[nodisplayskipstretch]{setspace}

\DeclareMathOperator*{\argmax}{argmax}
\DeclareMathOperator*{\argmin}{argmin}

\title{Bayesian Optimization for Personalized Dose-Finding Trials with Combination Therapies}

\author{
  James Willard, Shirin Golchi, Erica EM Moodie \\
  Department of Epidemiology and Biostatistics \\
  McGill University \\
  Montreal, Canada \\
  \And
  Bruno Boulanger, Bradley P. Carlin \\
  Cencora-PharmaLex \\
  Mont-Saint-Guibert, Belgium}

\begin{document}
\maketitle

\begin{abstract}
Identification of optimal dose combinations in early phase dose-finding trials is challenging, due to the trade-off between precisely estimating the many parameters required to flexibly model the possibly non-monotonic dose-response surface, and the small sample sizes in early phase trials. This difficulty is even more pertinent in the context of personalized dose-finding, where patient characteristics are used to identify tailored optimal dose combinations. To overcome these challenges, we propose the use of Bayesian optimization for finding optimal dose combinations in standard (``one size fits all") and personalized multi-agent dose-finding trials.  Bayesian optimization is a method for estimating the global optima of expensive-to-evaluate objective functions.  The objective function is approximated by a surrogate model, commonly a Gaussian process, paired with a sequential design strategy to select the next point via an acquisition function. This work is motivated by an industry-sponsored problem, where focus is on optimizing a dual-agent therapy in a setting featuring minimal toxicity. To compare the performance of the standard and personalized methods under this setting, simulation studies are performed for a variety of scenarios. Our study concludes that taking a personalized approach is highly beneficial in the presence of heterogeneity.
\end{abstract}


\maketitle

\newpage
\fontsize{12}{24}\selectfont

\section{Introduction}
Early phase clinical trials are designed to assess the safety and efficacy profiles of first in human doses of an experimental drug. Traditional adaptive dose-finding designs fall into three major families: algorithmic designs, model-assisted designs, and model-based designs \citep{berry2010bayesian}. Here, our interest lies in model-based designs, which have been extended to the dual-agent dose combination setting (e.g., \citealt{wang2005two, wages2014phase}). Some of these designs (e.g., \citealt{houede2010utility, wang2023bayesian}) propose using flexible models to handle possible non-monotonicities in the dose-efficacy/dose-toxicity surfaces, recognizing that monotonicity depends on the type of drug being administered and may not hold in general \citep{li2017toxicity}. Additionally, existing methods often restrict dose-finding to a small set of pre-selected dose combinations, which can fail to identify the optimal dose combination \citep{hirakawa2015comparative}. Adaptive dose insertion designs have been proposed to allow for the evaluation of additional dose combinations if warranted by the data \citep{cai2014bayesian, lyu2019aaa}. 

Most existing dose-finding designs, including those mentioned above, seek to find optimal dose combinations without consideration of additional covariate information (we refer to this as \textit{standard dose-finding}). These designs ignore the possibility that patient responses may be variable across different subgroups within the population. With recent advances in molecular biology, specifically the identification of novel biomarkers, interest has grown in personalized (or precision) medicine. \textit{Personalized dose-finding} aims to find optimal dose combinations based on individual patient characteristics. When utilizing parametric models, as in many of the standard dose-finding methods described above, extension to the personalized setting is challenging due to the limited sample sizes and potentially large number of dose-covariate interaction terms required. For single-agent personalized dose-finding, dimension-reduction methods have been proposed \citep{guo2022bayesian}. \cite{mozgunov2022dose} has considered dual-agent personalized dose-finding, but where the patient-specific dose of one of the agents is selected externally by clinicians. To our knowledge, there are currently no multi-agent personalized dose-finding designs where all dose dimensions are explored during optimization. 

Our work is motivated by a problem in industry, where a sponsor is interested in a design for the development of an intraocular implant that combines two topical agents.  Each agent has been in use and is well tolerated.  No drug-related adverse events are expected, and we anticipate minimal toxicity, if any. Interest lies in obtaining the optimal dose combination of these two agents using a continuous efficacy measure, which need not be monotonic with the doses of each agent. Additionally, response heterogeneity is expected to exist with respect to a key binary covariate, a particular characteristic of the lens of the eye, and we seek a design which accommodates this feature. We allow for exploration of the dose combination region, which has been defined using data from previous observational and randomized studies, without consideration of formal dose escalation/de-escalation rules. As each new dose combination investigated carries engineering costs, early stopping is desirable if warranted by the data.

Considering this motivating example, we propose novel methodology for multi-agent dose-finding which meets several criteria.  First, the model estimating the response surface is parsimonious, which is suitable for small sample sizes, yet remains flexible enough to model both monotonic and non-monotonic dose-response surfaces.  Next, the design sequentially explores the entire dose combination space in a principled way, and allows for early stopping if warranted.  Finally, our methodology facilitates extension to personalized dose-finding. Specifically, we propose modelling the response surface with a Gaussian process (GP) and using Bayesian optimization for efficient standard and personalized multi-agent dose-finding, assuming minimal toxicity. Bayesian optimization has been previously proposed for monotherapy in a standard dose-finding setting \citep{takahashi2021phase12, takahashi2021bayesian}. Our contributions here differ from this previous work by utilizing a different sequential search policy, by extending the approach to combination therapies and personalized dose-finding setting, and by permitting early stopping. 

The remainder of this manuscript is organized in the following manner. We first introduce Bayesian optimization, propose methods for standard and personalized multi-agent dose-finding, and describe an early stopping rule. A simulation study is then performed to compare the standard and personalized dose-finding methods for a dual-agent therapy under a variety of scenarios without early stopping. We then consider the development of the previously described intraocular implant and compare several dose-finding designs which include early stopping rules. We conclude with a discussion and possible directions for future work.

\section{Bayesian Optimization for Dose-Finding}\label{sec:bayesopt}
\textbf{Standard Dose-Finding.} Define $f^{\dagger}(\mathbf{d})$ to be a continuous dose-response surface of interest.  This may be the dose-efficacy surface or the dose-utility surface, where utility is defined using both efficacy and toxicity outcomes. For the remainder of the manuscript, we assume $f^{\dagger}(\mathbf{d})$ is the dose-efficacy surface, though we note that all proposed methods apply to the dose-utility setting. We assume minimal toxicity over the entire dose combination space, $\mathbb{D}$. Thus, dose escalation/de-escalation rules are not considered, and it is ethically permissible to select future dose combinations throughout the design space.  Our interest is in obtaining the combination of $J$ dosing agents $\mathbf{d}=(d_1,..., d_J) \in \mathbb{D} \subset \mathbb{R}^{J}$ which maximizes $f^{\dagger}(\mathbf{d})$.  That is, the search for the optimal dose combination is not restricted to a set of pre-defined doses, but rather a subspace of $\mathbb{R}^{J}$. Since optimization problems are commonly viewed as minimizing an objective function, the objective function $f(\mathbf{d})$ for $f^{\dagger}(\mathbf{d})$ can be defined as:
\begin{equation}
    f(\mathbf{d}) = \begin{cases}
    - f^{\dagger}(\mathbf{d}) & \text{if positive values of $f^{\dagger}(\mathbf{d})$ are preferred} \\
    f^{\dagger}(\mathbf{d}) & \text{otherwise.} \nonumber
    \end{cases}
\end{equation}
\noindent Then the goal is to identify the optimal dose combination:
\begin{equation}
    \argmin_{\mathbf{d} \in \mathbb{D}} f(\mathbf{d}). \nonumber
\end{equation}
To do so, we use Bayesian optimization which is a derivative-free optimization method that estimates the global optima of expensive-to-evaluate objective functions \citep{garnett2023bayesian}. It is a sequential design strategy that approximates the objective function with a stochastic surrogate model (commonly a GP), and selects the next point by optimizing an \textit{acquisition function}, $\alpha(\widetilde{\mathbf{d}} \mid \mathcal{D})$. Bayesian optimization is commonly used in engineering, machine learning, and computer experiments \citep{gramacy2020surrogates, murphy_pml2Book}, and has recently been employed to efficiently find optimal dynamic treatment regimes \citep{duque2022estimation, freeman2022dynamic}. In this work, we model the objective function with a GP, and select the next dose combination as that which maximizes an acquisition function. This process repeats until optimality criteria are satisfied or sample size limits are reached. 

To optimize $f(\mathbf{d})$, $r$ patients are assigned to each of $c$ initial dose combinations, yielding $n=r \times c$ patient responses which are treated as noisy observations $y_i = f(\mathbf{d}_i) + \epsilon_i$, where $\epsilon_i \sim N(0,\sigma^2_y)$ for $i=1,...,n$. A GP prior is then placed on the latent objective function,
\begin{equation}
    f(\mathbf{d}) \sim GP(m(\mathbf{d}), \mathcal{K}(\mathbf{d}, \mathbf{d}^{\prime})) \nonumber
\end{equation}
where $m(\mathbf{d}) = \mathbb{E}[f(\mathbf{d})]$ is the mean function, and $\mathcal{K}(\mathbf{d}, \mathbf{d}^{\prime})=\mathbb[(f(\mathbf{d}) - m(\mathbf{d}))(f(\mathbf{d}^{\prime}) - m(\mathbf{d}^{\prime}))^{T}]$ is a covariance function with a positive definite kernel \citep{williams2006gaussian}. Since the method's flexibility is largely determined by the choice of kernel function, constant mean GPs can be used. However, information about the response surface (e.g., pharmacokinetic/pharmacodynamic models) can be incorporated through the mean function if desired. We employ GP models with a constant mean function, $m(\mathbf{d}) = \beta_0$, and a separable anisotropic squared exponential kernel,
\begin{equation}
    \mathcal{K}(\mathbf{d},\mathbf{d}^{\prime})=\exp\left\{-\sum_{j=1}^{J}\frac{(d_j - d_j^{\prime})^2}{2l_{d_j}^2}\right\}. \label{squared_exp_anisotrop}
\end{equation}
This kernel is parameterized by characteristic length-scales $l_{d_j}$, which control the rate of decay in correlation between two dose combinations with respect to each dosing dimension. Additionally, it is an example of a stationary kernel, which assumes the degree of correlation between two dose combinations depends only on their distance from one another, not on their locations in the dose combination space. We assume stationary dose-response surfaces throughout this work, but comment on relaxing this assumption in the discussion. This induces a multivariate normal distribution on the observations, 
$\mathbf{y} \sim N(\beta_0 \mathbf{1}_n, \nu\mathbf{K})$, with positive definite covariance matrix, $\mathbf{K}(i,j)=\mathcal{K}(\mathbf{d}_i,\mathbf{d}_j) + \tau^2\mathds{1}_{i=j}$, that contains noise parameter $\tau^2$ and is multiplied by scale parameter $\nu$, which determines the variability of the objective function throughout the dose combination space. We note that $\sigma^2_y=\nu\tau^2$ but that the model is parameterized as above to facilitate faster inference (see section 5.2.2 of \cite{gramacy2020surrogates} for details). After observing data $\mathcal{D} = \{(\mathbf{d}_i,y_i)\}_{i=1}^{n}$, we adopt an empirical Bayes approach toward the kernel hyperparameters $\boldsymbol{\theta} = \{\nu, \tau^2, l_{d_1}, l_{d_2}\}$, replacing them with their maximum likelihood estimates \citep{williams2006gaussian, gramacy2020surrogates, hetgp}. The posterior distribution of $\mathbf{f}$ at a new dose combination $\widetilde{\mathbf{d}}$ is then $p(\mathbf{f} \mid \mathcal{D}, \widetilde{\mathbf{d}}) = N(\mu(\widetilde{\mathbf{d}}), \sigma^2(\widetilde{\mathbf{d}}))$  \citep{hetgp}, such that
\begin{equation}
\begin{aligned}
    \mu(\widetilde{\mathbf{d}}) &= \hat{\beta}_0 + \mathbf{k}(\widetilde{\mathbf{d}})^{T}\mathbf{K}^{-1}(\mathbf{y} - \hat{\beta}_0\mathbf{1})\\
    \sigma^2(\widetilde{\mathbf{d}}) &= \nu\mathcal{K}(\widetilde{\mathbf{d}},\widetilde{\mathbf{d}}) - \nu\mathbf{k}(\widetilde{\mathbf{d}})^{T}\mathbf{K}^{-1}\mathbf{k}(\widetilde{\mathbf{d}}) + \frac{(1 -\mathbf{k}(\widetilde{\mathbf{d}})^{T}\mathbf{K}^{-1}\mathbf{1})^2}{\mathbf{1}^T\mathbf{K}^{-1}\mathbf{1}} \\
    \hat{\beta}_0 &= \frac{\mathbf{1}^T\mathbf{K}^{-1}\mathbf{y}}{\mathbf{1}^T\mathbf{K}^{-1}\mathbf{1}} 
\end{aligned}
\end{equation}
where $\mathbf{K}$ is $n \times n$ and $\mathbf{k}(\widetilde{\mathbf{d}})=(\mathcal{K}(\mathbf{d}_1, \widetilde{\mathbf{d}}),..., \mathcal{K}(\mathbf{d}_n, \widetilde{\mathbf{d}}))^{T}$ is $n \times 1$.

The next dose combination for evaluation, denoted by $\mathbf{d}^{(c+1)}$, is then selected as that candidate dose combination $\widetilde{\mathbf{d}} \in \mathbb{D}$ which maximizes an acquisition function:
\begin{equation}
    \mathbf{d}^{(c+1)} = \argmax_{\widetilde{\mathbf{d}} \in \mathbb{D}} \alpha(\widetilde{\mathbf{d}} \mid \mathcal{D}). \nonumber
\end{equation}
\noindent One acquisition function commonly paired with a GP is the Expected Improvement (EI), defined as
\begin{equation}
    \alpha_{EI}(\widetilde{\mathbf{d}} \mid \mathcal{D}) = \mathbb{E}[\max(0,f^*-f(\widetilde{\mathbf{d}})) \mid \mathcal{D},\widetilde{\mathbf{d}}] \nonumber
\end{equation}  
and available in closed form \citep{jones1998efficient},
\begin{equation}
    \alpha_{EI}(\widetilde{\mathbf{d}} \mid \mathcal{D}) = (f^* - \mu(\widetilde{\mathbf{d}}))\Phi\left(\frac{f^*-\mu(\widetilde{\mathbf{d}})}{\sigma({\widetilde{\mathbf{d}})}}\right) + \sigma(\widetilde{\mathbf{d}})\phi\left(\frac{f^*-\mu(\widetilde{\mathbf{d}})}{\sigma({\widetilde{\mathbf{d}})}}\right), \nonumber
\end{equation}
where $f^*$ denotes the current optimum, $\Phi(\cdot)$ and $\phi(\cdot)$ denote the cdf and pdf of a standard normal random variable, respectively, and where $\mu(\widetilde{\mathbf{d}})$ and $\sigma(\widetilde{\mathbf{d}})$ are the posterior mean and standard deviation of $f$ evaluated at $\widetilde{\mathbf{d}}$, respectively.  The EI balances between \textit{exploiting} regions that have desirable values of $f(\mathbf{d})$ (first term), and \textit{exploring} regions in the dose combination space that are imprecisely estimated (second term).  Under a noisy setting, we can set $f^*=\min_{\mathbf{d}}\widetilde{\boldsymbol{\mu}}$, the minimum of the posterior mean \citep{gramacy_lee_2011, picheny2013quantile}. A different acquisition function, called  the Augmented Expected Improvement (AEI; \cite{huang2006global}), has been shown to offer better performance than EI under higher noise settings \citep{picheny2013benchmark}. AEI defines $f^* = \mu(\mathbf{d}^{*})$, the posterior mean of $f$ at the current ``effective best solution", $\mathbf{d}^{*}$, which can be defined as the point which minimizes a posterior $\beta$-quantile. We follow the recommendation in \cite{huang2006global} and define $\mathbf{d}^{*}$ as that point which minimizes the $\beta=0.84$ posterior quantile, which is equivalent to setting $\mathbf{d}^{*}= \argmin_{\widetilde{\mathbf{d}}} \mu(\widetilde{\mathbf{d}}) + \gamma\sigma(\widetilde{\mathbf{d}})$ when $\gamma=1$. The AEI is then defined as:
\begin{equation}
    \alpha_{AEI}(\widetilde{\mathbf{d}} \mid \mathcal{D}) = \alpha_{EI}(\widetilde{\mathbf{d}} \mid \mathcal{D})\left(1 - \frac{\tau}{\sqrt{\sigma^2(\widetilde{\mathbf{d}}) + \tau^2}}\right).
\end{equation}
The multiplicative term serves to promote exploration by penalizing dose combinations which have small posterior variance \citep{picheny2013benchmark}. In this work, we use AEI since we find it offers moderate improvement over EI.

After evaluation of $\mathbf{d}^{(c+1)}$ in $r$ new patients, the data are updated, $\mathcal{D}=\mathcal{D} \cup \{(\mathbf{d}^{(c+1)}_i,y_i)\}_{i=1}^{r}$. The GP model is refit to obtain a new posterior distribution $p(\mathbf{f} \mid \mathcal{D}, \widetilde{\mathbf{d}})$.  Then $s=1,...,S$ samples $f_s(\widetilde{\mathbf{d}})$ from this posterior are obtained to yield $S$ samples from the posterior of the optimal dose combination $p(\mathbf{d}_{opt} \mid \mathcal{D})$ as:
\begin{equation}
    \mathbf{d}_{opt,s} = \argmin_{\widetilde{\mathbf{d}} \in \mathbb{D}} f_s(\widetilde{\mathbf{d}}). \nonumber
\end{equation}
This procedure continues until the sample size limit is reached or an early stopping rule, which we denote by $\mathds{1}_{\text{STOP}}$, is satisfied. 

One possible early stopping rule proposed in \cite{huang2006global} is to allow stopping only after $\max_{\widetilde{\mathbf{d}} \in \mathbb{D}} \alpha_{AEI}(\widetilde{\mathbf{d}} \mid \mathcal{D}) < \delta$. In this case, the algorithm terminates only if there is little improvement to be gained over $f^*$ across the dose combination space. Under a noisy setting, the authors suggest this be satisfied for $(J+1)$ consecutive algorithm iterations before termination. This yields the following stopping rule:
\begin{equation}
\mathds{1}_{\text{STOP}} = 
\begin{cases}
    \text{TRUE} & \max_{\widetilde{\mathbf{d}} \in \mathbb{D}} \alpha_{AEI}(\widetilde{\mathbf{d}} \mid \mathcal{D}) < \delta\text{ for }(J+1)\text{ iterations} \\
    \text{FALSE} & \text{otherwise.}
\end{cases} \label{stopping_rule}
\end{equation}
We note that $\delta$ is a tuning parameter that controls performance of the algorithms. Its value can be determined through sensitivity analysis using several values obtained through Monte Carlo simulation. For example, the Monte Carlo distributions of $\alpha_{AEI}$ can be obtained at each iteration and different values of $\delta$ can be selected as summary statistics of these distributions (e.g., the median). The performance of these values can then be compared, with smaller values of $\delta$ implying more stringent stopping criteria, and larger values of $\delta$ permitting earlier stopping.

The sequential procedure and stopping rule described above suggest Algorithm \ref{alg:standard}, referred to as the \textit{standard optimization algorithm}. In many applications of Bayesian optimization, it is standard to collect the initial data and then continue with $r=1$ at each iteration of the algorithm. This permits maximal exploration of the domain, as a novel input point from the design space can be evaluated each time.  In certain dose-finding applications, such as the motivating example of this work, engineering costs may prohibit the creation of a novel dose combination for each new patient, and some patients may necessarily be assigned to the same dose. 

\begin{algorithm}[t!]
\caption{Standard Optimization Algorithm}\label{alg:standard}
\begin{algorithmic}[1]
\Require $r$ patient responses at each of $c$ initial dose combinations, $\mathcal{D}=\{(\mathbf{d}_i, y_i)\}_{i=1}^{n_0}$ 
\State $n \gets n_0 = r \times c$ 
\State $\mathds{1}_{\text{STOP}} \gets \text{FALSE}$
\State Obtain $p(\mathbf{f} \mid \mathcal{D}, \widetilde{\mathbf{d}})$ and $p(\mathbf{d}_{opt} \mid \mathcal{D})$ using fitted GP \Comment{\parbox[t]{.3\linewidth}{Obtain posteriors}}
\State $\mathbf{d}^{*} \gets \argmin_{\widetilde{\mathbf{d}}} \mu(\widetilde{\mathbf{d}}) + \sigma(\widetilde{\mathbf{d}})$ \Comment{\parbox[t]{.3\linewidth}{Obtain effective best point}}
\State $f^* \gets \mu(\mathbf{d}^{*})$ \Comment{\parbox[t]{.3\linewidth}{Obtain best value of $f$}}
\State Calculate $\alpha_{AEI}(\widetilde{\mathbf{d}} \mid \mathcal{D})$ for $\widetilde{\mathbf{d}} \in \mathbb{D}$ \Comment{\parbox[t]{.3\linewidth}{Compute AEI}}
\While{$n<N$ and $\mathds{1}_{\text{STOP}} = \text{FALSE}$} 
    \State $\mathbf{d}^{(c+1)} \gets \argmax_{\widetilde{\mathbf{d}} \in \mathbb{D}} \alpha_{AEI}(\widetilde{\mathbf{d}} \mid \mathcal{D})$ \Comment{\parbox[t]{.3\linewidth}{Obtain next dose}}
    \For{$i=1,...,r$}
    \State Evaluate $y_i$ at $\mathbf{d}^{(c+1)}$ \Comment{\parbox[t]{.3\linewidth}{Observe outcomes}}
    \EndFor
    \State $n \gets n + r$; $c \gets c+1$  \Comment{\parbox[t]{.3\linewidth}{Update $n,c$}}
    \State $\mathcal{D} \gets \mathcal{D} \cup \{(\mathbf{d}^{(c+1)}_i,y_i)\}_{i=1}^{r}$ \Comment{\parbox[t]{.3\linewidth}{Update data}}
    \State Obtain $p(\mathbf{f} \mid \mathcal{D}, \widetilde{\mathbf{d}})$ and $p(\mathbf{d}_{opt} \mid \mathcal{D})$ using fitted GP \Comment{\parbox[t]{.3\linewidth}{Obtain posteriors}}
    \State $\mathbf{d}^{*} \gets \argmin_{\widetilde{\mathbf{d}}} \mu(\widetilde{\mathbf{d}}) + \sigma(\widetilde{\mathbf{d}})$ \Comment{\parbox[t]{.3\linewidth}{Obtain effective best point}}
    \State $f^* \gets \mu(\mathbf{d}^{*})$ \Comment{\parbox[t]{.3\linewidth}{Obtain best value of $f$}}
    \State Calculate $\alpha_{AEI}(\widetilde{\mathbf{d}} \mid \mathcal{D})$ for $\widetilde{\mathbf{d}} \in \mathbb{D}$ \Comment{\parbox[t]{.3\linewidth}{Compute AEI}}
    \State Update $\mathds{1}_{\text{STOP}}$ using (\ref{stopping_rule}) \Comment{\parbox[t]{.3\linewidth}{Update stopping rule}}
\EndWhile
\end{algorithmic}
\end{algorithm}

\textbf{Personalized Dose-Finding.} Personalized medicine recognizes that response heterogeneity may exist within the population of interest; personalized dose-finding incorporates covariate information to account for this. Consider a set of $P$ discrete covariates $\mathbf{Z}=\{Z_p\}_{p=1}^P$.  The Cartesian product of the levels of these $P$ covariates define $K$ strata. Personalized dose finding seeks to find the optimal dose combinations across the continuous dose combination space for each of the $K$ strata:
\begin{equation}
    \argmin_{\mathbf{d} \in \mathbb{D}} f(\mathbf{d},\mathbf{z}_k)\ \ \text{ for }\ \ k=1,...,K. \nonumber
\end{equation}
The objective function $f(\mathbf{d},\mathbf{z})$ is modeled using a single GP fit to the data $\mathcal{D}=\{(\mathbf{d}_i,\mathbf{z}_i, y_i)\}_{i=1}^n$, where $y_i = f(\mathbf{d}_i,\mathbf{z}_i) + \epsilon_i$ with $\epsilon_i \sim N(0,\sigma^2_y)$, which allows information to be borrowed across strata.  One possible method of incorporating the additional covariates into the GP model is through the kernel function.  We use a (stationary) separable anisotropic squared exponential kernel function that includes the additional covariates of interest:
\begin{equation}\label{cov_squared_exp_kernel}
     \mathcal{K}((\mathbf{d},\mathbf{z}),(\mathbf{d},\mathbf{z})^{\prime})=\exp\left\{-\left(\sum_{j=1}^{J}\frac{(d_j - d_j^{\prime})^2}{2l_{d_j}^2} + \sum_{p=1}^{P}\frac{(z_p - z_p^{\prime})^2}{2l_{p}^2}\right)\right\}. 
\end{equation}
As before, we use an empirical Bayes approach for estimating the hyperparameters. In the case where $Z_p$ is a binary variable representing the levels of covariate $p$, the correlation between two patient responses is reduced by a factor of  $\exp(-1/(2l^2_p))$ if they belong to different strata with respect to covariate $p$. 

Since the objective function may exhibit different behavior within each stratum, each stratum may have a unique best objective function value $f_k^*$. Similar to the standard case, $f_k^*$ is estimated as the posterior mean at the effective best point within stratum $k$, $f_k^*= \mu(\mathbf{d}^{*},\mathbf{z}_k)$. To account for this heterogeneity, the sequential selection is performed within each stratum but with the GP fit utilizing data from all strata. This proceeds by modifying the AEI acquisition function to use $f_k^*$ when conditioned on being in stratum $k$, denoted by $\alpha_{AEI}(\widetilde{\mathbf{d}} \mid \mathcal{D}, \widetilde{\mathbf{z}}_k)$. The sequential procedure continues until sample size limits are reached or until an early stopping rule is satisfied for each stratum. Since optimal doses in some strata may be easier to identify than in others, stratum-specific early stopping should be employed. One possible stratum specific stopping rule, which we denote by $\mathds{1}_{\text{STOP},k}$, replaces $\alpha_{AEI}(\widetilde{\mathbf{d}} \mid \mathcal{D})$ in (\ref{stopping_rule}) with $\alpha_{AEI}(\widetilde{\mathbf{d}} \mid \mathcal{D}, \widetilde{\mathbf{z}}_k)$. Upon termination of the algorithm, the posterior distribution of the optimal dose combination for each strata is returned, denoted by $p(\mathbf{d}_{opt} \mid \mathcal{D}, \mathbf{z}_k)$. These modifications suggest Algorithm \ref{alg:personalized}, the \textit{personalized optimization algorithm}.

\begin{algorithm}[t!]
\caption{Personalized Optimization Algorithm}\label{alg:personalized}
\begin{algorithmic}[1]
\Require $r$ patient responses at each of $c_k$ initial dose combinations per stratum, ${\mathcal{D}=\{(\mathbf{d}_i,\mathbf{z}_i, y_i)\}_{i=1}^{n_0}}$
\State $n \gets n_0=\sum_{k=1}^Kn_{0,k}$ 
\State Obtain $p(\mathbf{f} \mid \mathcal{D}, \widetilde{\mathbf{d}},  \widetilde{\mathbf{z}})$ using fitted GP \Comment{\parbox[t]{.3\linewidth}{Obtain posterior of $f$}}
\For{$k=1,...,K$}
    \State Obtain $p(\mathbf{d}_{opt} \mid \mathcal{D}, \mathbf{z}_k)$ \Comment{\parbox[t]{.3\linewidth}{Obtain posterior of $\mathbf{d}_{opt}$}}
    \State $\mathbf{d}^{*} \gets \argmin_{\widetilde{\mathbf{d}}} \mu(\widetilde{\mathbf{d}}, \mathbf{z}_k) + \sigma(\widetilde{\mathbf{d}}, \mathbf{z}_k)$ \Comment{\parbox[t]{.3\linewidth}{Obtain effective best point}}
    \State $f^*_k \gets \mu(\mathbf{d}^{*}, \mathbf{z}_k)$ \Comment{\parbox[t]{.3\linewidth}{Obtain best value $f^*_k$}}
    \State Calculate $\alpha_{AEI}(\widetilde{\mathbf{d}} \mid \mathcal{D}, \widetilde{\mathbf{z}}_k)$ for $\widetilde{\mathbf{d}} \in \mathbb{D}$ \Comment{\parbox[t]{.3\linewidth}{Compute AEI}}
    \State $\mathds{1}_{\text{STOP},k} \gets \text{FALSE}$
\EndFor
\While{$n<N$ and $\mathds{1}_{\text{STOP},k} = \text{FALSE}$ for at least one stratum $k$}
    \For{$k=1,...,K$}
    \State $n_k \gets 0$
    \State $\mathcal{D}_k = \emptyset$
    \If{$\mathds{1}_{\text{STOP},k} = \text{FALSE}$}
        \State $\mathbf{d}^{(c_k+1)} \gets \argmax_{\tilde{\mathbf{d}} \in \mathbb{D}} \alpha_{AEI}(\tilde{\mathbf{d}} \mid \mathcal{D}, \widetilde{\mathbf{z}}_k)$ \Comment{\parbox[t]{.3\linewidth}{Obtain next dose}}
        \For{$i=1,...,r$}
        \State Evaluate $y_i$ at $(\mathbf{d}^{(c_k+1)},\mathbf{z}_k)$ \Comment{\parbox[t]{.3\linewidth}{Observe outcomes}}
        \EndFor
        \State $n_k \gets r$; $c_k \gets c_k+1$ \Comment{\parbox[t]{.3\linewidth}{Update $n_k, c_k$}}
        \State $\mathcal{D}_k = \{(\mathbf{d}^{(c_k+1)}_i,\mathbf{z}_k,y_i)\}_{i=1}^r$
    \EndIf
    \EndFor
    \State $n \gets n + \sum_{k=1}^Kn_k$ \Comment{\parbox[t]{.3\linewidth}{Update $n$}}
    \State $\mathcal{D} = \mathcal{D} \bigcup_{k=1}^{K}\mathcal{D}_k$ \Comment{\parbox[t]{.3\linewidth}{Update data}}
    \State Obtain $p(\mathbf{f} \mid \mathcal{D}, \widetilde{\mathbf{d}},  \widetilde{\mathbf{z}})$ using fitted GP \Comment{\parbox[t]{.3\linewidth}{Obtain posterior of $f$}}
    \For{$k=1,...,K$}
    \State Obtain $p(\mathbf{d}_{opt} \mid \mathcal{D}, \mathbf{z}_k)$ \Comment{\parbox[t]{.3\linewidth}{Obtain posterior of $\mathbf{d}_{opt}$}}
    \If{$\mathds{1}_{\text{STOP},k} = \text{FALSE}$}
        \State $\mathbf{d}^{*} \gets \argmin_{\widetilde{\mathbf{d}}} \mu(\widetilde{\mathbf{d}}, \mathbf{z}_k) + \sigma(\widetilde{\mathbf{d}}, \mathbf{z}_k)$ \Comment{\parbox[t]{.3\linewidth}{Obtain effective best point}}
        \State $f^*_k \gets \mu(\mathbf{d}^{*}, \mathbf{z}_k)$ \Comment{\parbox[t]{.3\linewidth}{Obtain best value $f^*_k$}}
        \State Calculate $\alpha_{AEI}(\widetilde{\mathbf{d}} \mid \mathcal{D}, \widetilde{\mathbf{z}}_k)$ for $\widetilde{\mathbf{d}} \in \mathbb{D}$ \Comment{\parbox[t]{.3\linewidth}{Compute AEI}}
        \State Update $\mathds{1}_{\text{STOP},k}$ using (\ref{stopping_rule}) \Comment{\parbox[t]{.3\linewidth}{Update stopping rule}}
    \EndIf
    \EndFor
\EndWhile   
\end{algorithmic}
\end{algorithm}

\section{Simulation Study} \label{sec:sim_study}

Below we perform a simulation study to compare the performance of the standard and personalized algorithms (Algorithms \ref{alg:standard} and \ref{alg:personalized}, respectively) under three scenarios with no early stopping (i.e., $\delta=0$ in (\ref{stopping_rule})). Early stopping will be investigated in the next section. Scenarios 1 and 2 consider a single binary covariate $Z_1$. We index the true optimal dose combinations, $\mathbf{d}_{opt,k}$, and the true optimal values of the objective function, $f_{opt,k}$, using the values of $Z_1$.  That is, when $Z_1=0$ we use $\mathbf{d}_{opt,0}$ and $f_{opt,0}$. Scenario 3 considers two binary covariates, $Z_1$ and $Z_2$, and the $\mathbf{d}_{opt,k}$ and $f_{opt,k}$ are indexed similarly. For example, when $Z_1=0$ and $Z_2=1$, we use $\mathbf{d}_{opt,01}$ and  $f_{opt,01}$. To make the simulations in this manuscript more computationally feasible, we modify Algorithms \ref{alg:standard} and \ref{alg:personalized} to return a point estimate of $\mathbf{d}_{opt,k}$ rather than the entire posterior distribution. The point estimate is defined as the minimizer of the posterior mean surface, $\widehat{\mathbf{d}}_{opt,k} = \argmin_{\widetilde{\mathbf{d}}}\mu(\widetilde{\mathbf{d}}, \widetilde{\mathbf{z}}_k)$.

We utilize dose combinations $\mathbf{d}=(d_1,d_2) \in [0,1]^2$, assumed to be standardized, where $\mathbf{d} = (0,0)$ corresponds to the combination using the lowest doses of interest for each agent and where $\mathbf{d}=(1,1)$ corresponds to the combination using the highest doses of interest for each agent. The point estimates, $\widehat{\mathbf{d}}_{opt,k}$, and the next dose combinations for evaluation, $\mathbf{d}^{(c_k+1)}$, are set, respectively, as the minimizers of $\widetilde{\boldsymbol{\mu}}(\mathbf{d}, \mathbf{z}_k)$ and maximizers of $\alpha_{AEI}(\widetilde{\mathbf{d}} \mid \mathcal{D}, \widetilde{\mathbf{z}}_k)$. The $\alpha_{AEI}(\widetilde{\mathbf{d}} \mid \mathcal{D}, \widetilde{\mathbf{z}}_k)$ is evaluated across an evenly spaced grid on $[0,1]^2$. The grid is incremented by $0.25$ in each dimension, reflecting the degree of precision to which the drug maker can manufacture a particular dose combination.  As a result, some dose combinations may be suggested more than once in the algorithm. While the proposed method is capable of optimizing over the continuous dose combination space, it is important to incorporate any manufacturing constraints into the optimization procedures to avoid suggesting doses which are not feasible to engineer. 

\begin{figure}
    \includegraphics[width=\linewidth]{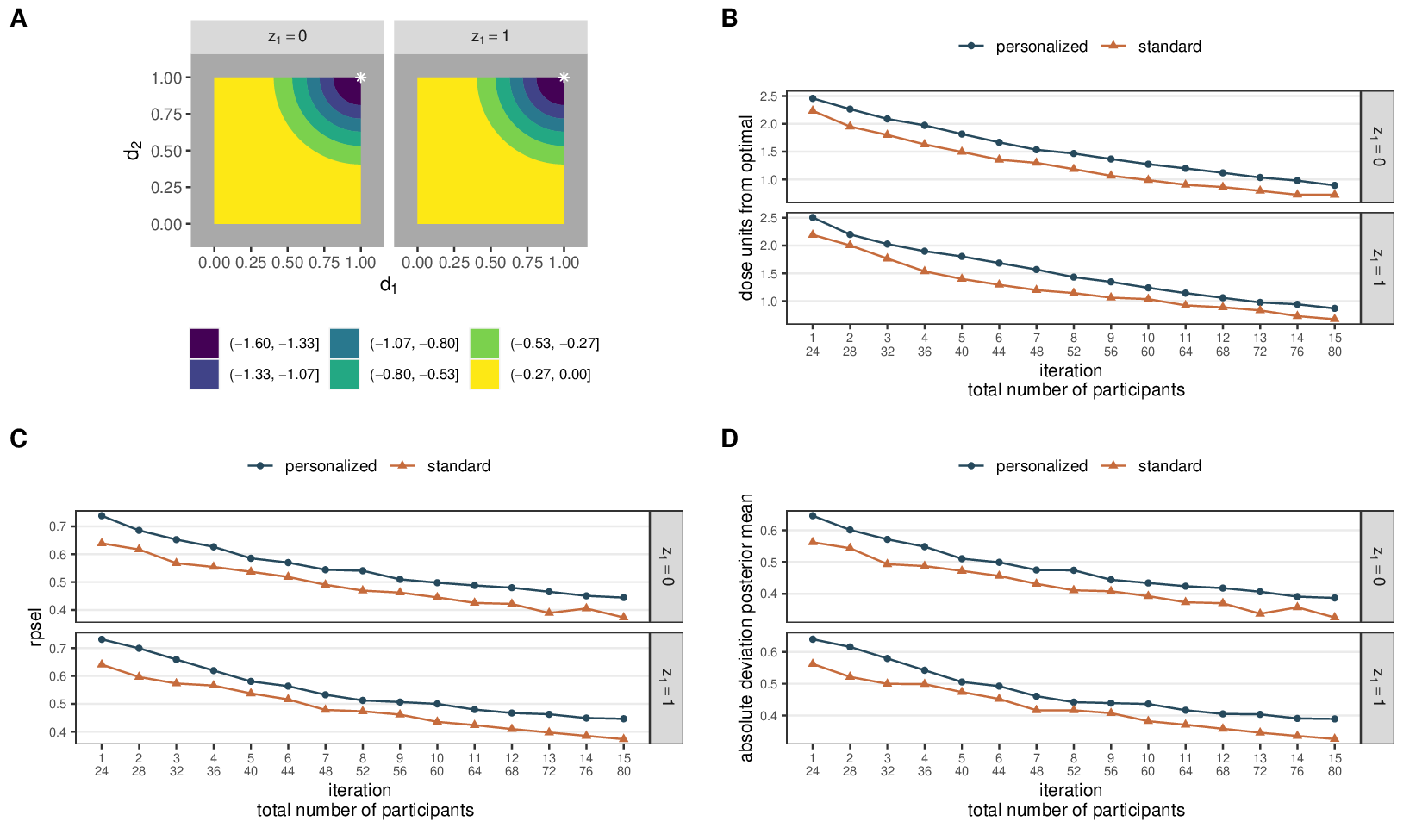}
    \caption{Scenario 1. A) Objective function, white stars denote $\mathbf{d}_{opt,k}$, B) expected dose units from the optimal dose combination as defined in (\ref{dose_unit_distance}) by iteration, C) average RPSEL as defined in (\ref{rpsel}) by iteration, and D) average absolute deviation of the posterior mean estimate by iteration.}
    \label{fig:sim_scenario_1}
\end{figure}

\begin{figure}
    \includegraphics[width=\linewidth]{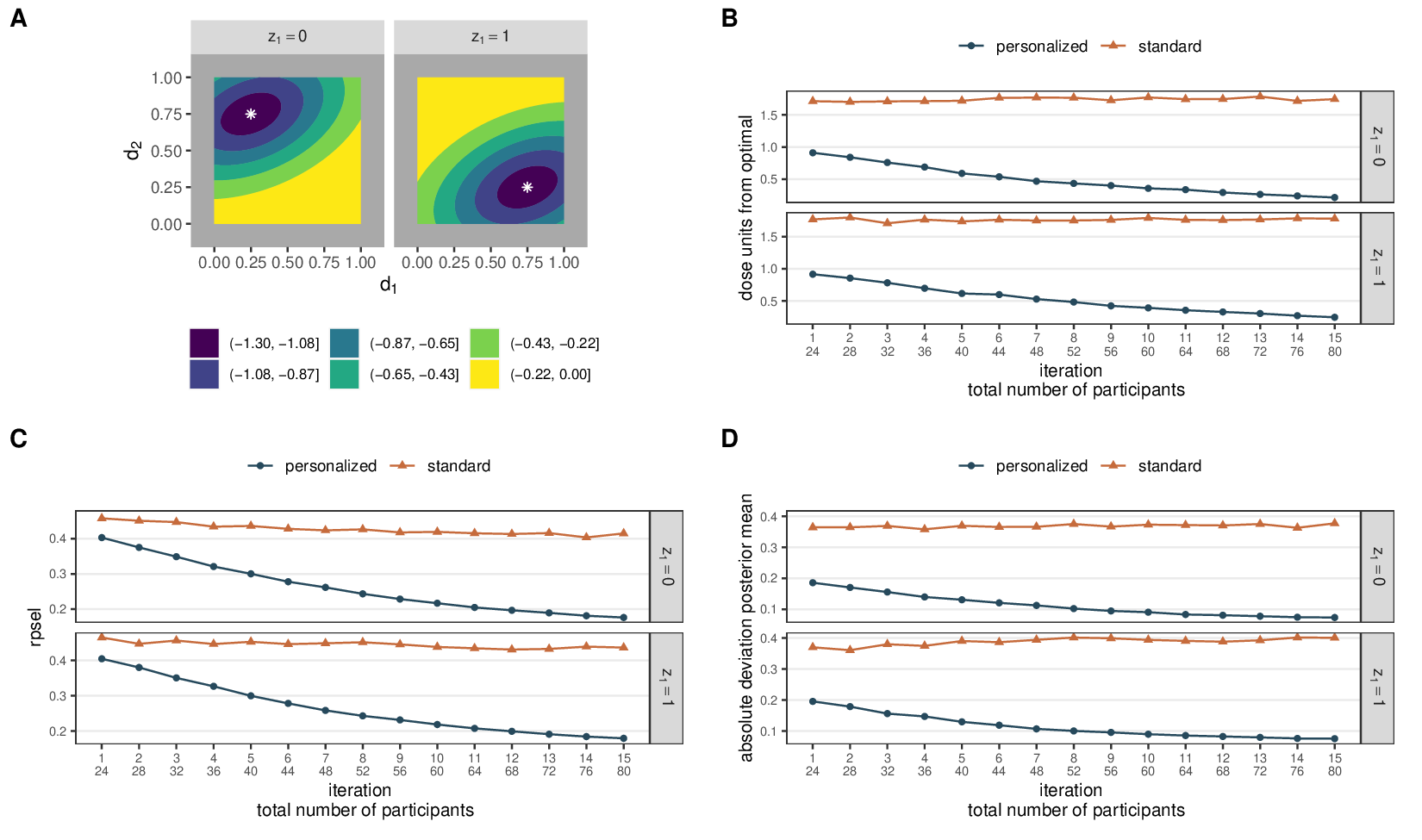}
    \caption{Scenario 2. A) Objective function, white stars denote $\mathbf{d}_{opt,k}$, B) expected dose units from the optimal dose combination as defined in (\ref{dose_unit_distance}) by iteration, C) average RPSEL as defined in (\ref{rpsel}) by iteration, and D) average absolute deviation of the posterior mean estimate by iteration.}
    \label{fig:sim_scenario_2}
\end{figure}

\begin{figure}
    \includegraphics[width=\linewidth]{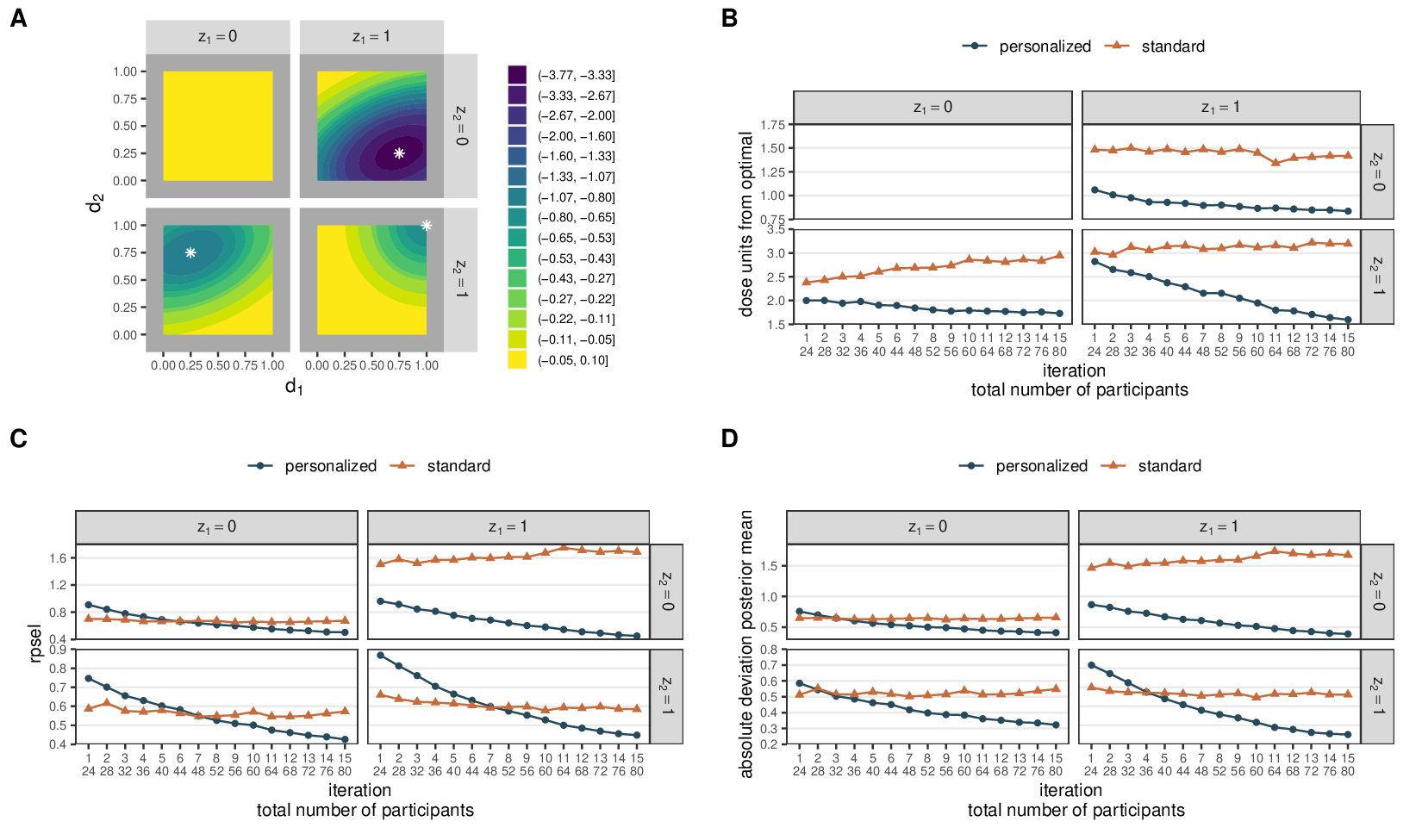}
    \caption{Scenario 3. A) Objective function, white stars denote $\mathbf{d}_{opt,k}$, B) expected dose units from the optimal dose combination as defined in (\ref{dose_unit_distance}) by iteration, C) average RPSEL as defined in (\ref{rpsel}) by iteration, and D) average absolute deviation of the posterior mean estimate by iteration. In (B), the plot for $(z_1=0,z_2=0)$ is empty as no optimal dose combination exists.}
    \label{fig:sim_scenario_3}
\end{figure}

As before, we define $f^{\dagger}(\mathbf{d},\mathbf{z})$ to be a continuous efficacy surface and assume we are in a minimal toxicity setting where it is ethically permissible to select future doses anywhere in the dose combination space. We define our objective function as the negative of the efficacy surface, $f(\mathbf{d},\mathbf{z})=-f^{\dagger}(\mathbf{d},\mathbf{z})$. The objective functions under the considered scenarios use the following negative multivariate normal densities:
\begin{equation}
\small
\begin{split}
    g_1(\mathbf{d}) =
        -N\begin{bmatrix}
        \begin{pmatrix}
        1\\
        1
        \end{pmatrix} ,
        \begin{pmatrix}
        0.1 & 0\\
        0 & 0.1
        \end{pmatrix}
        \end{bmatrix} 
\end{split}
\hspace{2mm}
\begin{split}
    g_2(\mathbf{d}) &= -N\begin{bmatrix}
        \begin{pmatrix}
        0.25\\
        0.75
        \end{pmatrix} ,
        \begin{pmatrix}
        0.2 & 0.05\\
        0.05 & 0.1
        \end{pmatrix}
        \end{bmatrix}
\end{split}
\hspace{2mm}
\begin{split}
        g_3(\mathbf{d}) &= 
        -N\begin{bmatrix}
        \begin{pmatrix}
        0.75\\
        0.25
        \end{pmatrix} ,
        \begin{pmatrix}
        0.2 & 0.05\\
        0.05 & 0.1
        \end{pmatrix}
        \end{bmatrix}.\nonumber
\end{split}
\end{equation}
The data generating mechanism for each scenario is $y = f(\mathbf{d},\mathbf{z}) + \epsilon$ where $\epsilon \sim N(0, \sigma^2_y)$, with the specification of $f(\mathbf{d},\mathbf{z})$ included in the first panel of Table \ref{tbl:sim_study_scenarios} (rows labeled ``Simulation Study") and plotted in panel A of Figures \ref{fig:sim_scenario_1}-\ref{fig:sim_scenario_3}. The values of $\sigma_y$ are chosen to ensure specific standardized effect sizes, defined as $ses=|f_{opt}|/\sigma_y$. We consider several standardized effect sizes drawn from a meta-analysis of dose-responses for a large drug development portfolio at a pharmaceutical company \citep{thomas2014meta}. We focus on standardized effect sizes for drugs that had laboratory confirmed endpoints, which is the type of endpoint used in our motivating example. The $25^{th}/50^{th}/75^{th}$ percentiles of these standardized effect sizes are $0.79/1/3.77$, which we refer to as small/medium/large effect sizes.

\begin{table}
    \caption{Simulation scenarios considered.  The data generating mechanism for each scenario is $y = f(\mathbf{d},\mathbf{z}) + \epsilon$ where $\epsilon \sim N(0, \sigma^2_y)$. The table columns contain the location of the optimal dose combination ($\mathbf{d}_{opt}$), the optimal value of the objective function ($f_{opt}$), the standardized effect size (ses), and whether or not the dose-efficacy surface is monotonically increasing with respect to each dosing dimension (monotone).}
    \centering
    \resizebox{\linewidth}{!}{
   \begin{tabular}[width=\linewidth]{clllllllc}
    \toprule
    & $f(\mathbf{d},\mathbf{z})$ & $\sigma_y$ & $z_1$ & $z_2$ & $\mathbf{d}_{opt}$ & $f_{opt}$ & ses & monotone \\
    \midrule 
    \raisebox{-0.95\height }{\multirow{4}{*}{\rotatebox[origin=r]{90}{{Simulation Study}}}}
    & 1) \multirow{2}{*}{\parbox{0.35\linewidth}{$\mathds{1}\{z_1=0\} \times g_1(\mathbf{d}) + \\  \mathds{1}\{z_1=1\} \times g_1(\mathbf{d})$}} & 2.015 & 0 & & $(1, 1)$ & -1.592 & 0.79 & Yes \\
    & & & 1 & & $(1, 1)$ &  -1.592 & 0.79 & Yes \\
    \cmidrule{2-9}
    & 2) \multirow{2}{*}{\parbox{0.35\linewidth}{$\mathds{1}\{z_1=0\} \times g_2(\mathbf{d}) + \\ \mathds{1}\{z_1=1\} \times g_3(\mathbf{d})$}}  & 0.319 & 0 & & $(0.25, 0.75)$ & -1.203 & 3.77 & No\\
    & & & 1 & & $(0.75, 0.25)$ &  -1.203 & 3.77 & No\\
    \cmidrule{2-9} 
    & 3) \multirow{4}{*}{\parbox{0.35\linewidth}{$\mathds{1}\{(z_1=0, z_2=0)\} \times 0 + \\ \mathds{1}\{(z_1=0,z_2=1)\} \times 0.831g_2(\mathbf{d}) +$ \\ 
    $\mathds{1}\{(z_1=1, z_2=0)\} \times 3.134g_3(\mathbf{d}) + \\ 
    \mathds{1}\{(z_1=1,z_2=1)\} \times 0.496g_1(\mathbf{d})$}} & 1 & 0 & 0 & None & None & 0 & No\\
    & & & 0 & 1 & $(0.25, 0.75)$ & -1 & 1 & No\\
    & & & 1 & 0 & $(0.75, 0.25)$ & -3.77 & 3.77 & No\\
    & & & 1 & 1 & $(1, 1)$ & -0.79 & 0.79 & Yes\\
    \midrule 
    \raisebox{-0.55\height}{\multirow{2}{*}{\rotatebox[origin=c]{90}{{Implant}}}}
    & 1) \multirow{2}{*}{\parbox{0.35\linewidth}{$\mathds{1}(z_1=0) \times 2.49g_2(\mathbf{d}) + \\ \mathds{1}(z_1=1) \times 6.65g_3(\mathbf{d}) - 2$}} & 5 & 0 & & $(0.25, 0.75)$ & -5 & 1 & No\\
    &  &  & 1 & & $(0.75, 0.25)$ & -10 & 2 & No \\ \vspace{1mm}
    \nonumber \\  
    \bottomrule
    \multicolumn{9}{l}{%
    \begin{minipage}{\linewidth}%
    \vspace{1mm}
    \footnotesize Note: The subtraction of $2$ in $f(\mathbf{d},\mathbf{z})$ under the Implant scenario corresponds to a base level of drug response outside the regions of optimality.%
    \end{minipage}%
    }
    \end{tabular}
    }
    \label{tbl:sim_study_scenarios}
\end{table}

Scenario 1 considers the case of no response heterogeneity across a binary covariate $Z_1$ and includes a small standardized effect size with a dose-efficacy surface which is monotonically increasing with respect to each dosing dimension. Both the locations of the optimal dose combinations, and the optimal values of the objective function, are the same across the strata. That is, $\mathbf{d}_{opt,0}=\mathbf{d}_{opt,1}$ and $f_{opt,0}=f_{opt,1}$.  Scenario 2 considers response heterogeneity across $Z_1$, where the locations of the optimal dose combinations differ. Under this scenario, $\mathbf{d}_{opt,0} \ne \mathbf{d}_{opt,1}$ but $f_{opt,0}=f_{opt,1}$. This scenario considers large standardized effect sizes with efficacy surfaces which are non-monotone with respect to each dosing dimension. Scenario 3 considers heterogeneity across two binary covariates, $Z_1$ and $Z_2$, where both the locations of the optimal dose combinations and the optimal values of the objective function differ across the strata. Thus, $\mathbf{d}_{opt,ij} \ne \mathbf{d}_{opt,lm}$ for $ij \ne lm$ and $f_{opt,ij} \ne f_{opt,lm}$ for $ij \ne lm$. This scenario includes a zero stratum, $(z_1=0,z_2=0)$, which represents the covariate pattern of those who do not respond to the drug. We note that in this stratum, $\mathbf{d}_{opt,00}$ and $f_{opt,00}$ do not exist, but we consider the standardized effect size to be 0. This scenario includes small, medium, and large standardized effect sizes as well as both monotone and non-monotone dose-efficacy surfaces.

For each scenario, the standard and personalized algorithms are run using a maximum sample size of 80 participants. While this number is larger than many early phase trials might be in practice, our goal is to investigate the algorithms' performance characteristics as the sample size increases. We defer the use of early stopping rules to the following section, but note that these will permit a reduction in the expected sample size. We assign participants to dose combinations such that the total sample size of each algorithm is equal at each iteration, which allows a comparison of their performance. For the standard algorithm under scenarios 1 and 2, $r=4$ participants are assigned to each dose combination on an initial dose matrix comprised of $c=5$ dose combinations, which are selected via a two-dimensional quasi-random Sobol sequence \citep{spacefillr}. This sequence serves as a space filling design and seeks to spread out the initial dose combinations in a more uniform manner than is typically accomplished via random sampling. More than one patient is assigned per dose combination to control the cost associated with producing novel dose-combinations, a financial constraint from our motivating problem. This yields $n_0=20$.  At each iteration of the algorithm, $r=4$ additional participants are assigned to each proposed dose combination.  This yields a total sample size of 80 after 15 iterations. We note that $r$ is a tuning parameter, and that reducing it will lead to more unique doses being explored for the same fixed sample size.  We explore this in the next section. 

For the personalized algorithm, $c=5$ initial dose combinations are selected in the same manner as above. To achieve the same total sample size by iteration as the standard algorithm, only $r=2$ participants are assigned to each dose combination within each stratum. This yields $n_{0,0}=n_{0,1}=10$, and so $n_0=20$, as in the standard algorithm.  At each iteration within each stratum, $r=2$ additional participants are assigned to each proposed dose combination. This yields a total sample size of 80 after 15 iterations. The same setup is used for scenario 3.  However, since the number of strata is doubled, the number of participants evaluated at each dose combination is halved for the personalized algorithm. Thus, the standard algorithm still evaluates $r=4$ participants per dose combination, but the personalized algorithm evaluates $r=1$ participant per dose combination, yielding a total sample size of 80 after 15 iterations. All computing is performed in the statistical programming language R \citep{r_language}. The objective functions are modeled using constant mean GP models which utilize the anisotropic separable squared exponential kernels previously described, with the hyperparameters being jointly optimized by maximizing the marginal log-likelihood of the data (i.e., empirical Bayes GP; \citealt{hetgp}). 

Algorithm performance is compared using several criteria which are estimated via $m=1,...,1000$ Monte Carlo simulations. The expected number of dosing units from the optimal dose combination is used to assess how close the recommended dose combination $\widehat{\mathbf{d}}_k$ is to $\mathbf{d}_{opt,k}$ at each iteration. This measure is defined as the expected value of the Euclidean distance between $\widehat{\mathbf{d}}_k$ and $\mathbf{d}_{opt,k}$ divided by the precision to which the sponsor can manufacture doses, which is $0.25$ in our simulations:
\begin{equation}\label{dose_unit_distance}
    E_{y \mid z_k}[\text{dose units}] \approx \frac{1}{1000} \sum_{m=1}^{1000} \frac{\sqrt{(\hat{d}^{(m)}_{1}-d_{1,opt})^2_k + (\hat{d}^{(m)}_{2}-d_{2,opt})^2_k}}{0.25}.
\end{equation}
We utilize the average root posterior squared error loss (RPSEL) to assess how well the true objective function value $f(\widehat{\mathbf{d}}_k, z_k)$ at the recommended dose is estimated by the pointwise posterior distribution of the objective function at the recommended dose, $p(f_k \mid \mathcal{D}, \widehat{\mathbf{d}}_{k}, z_k)$, where $f^{(s)}(\widehat{\mathbf{d}}_{k}, z_k)$ denotes a single posterior sample out of $s=1,...,10000$ posterior samples:
\begin{equation}\label{rpsel}
    \text{Average RPSEL} \approx \frac{1}{1000} \sum_{m=1}^{1000} \left[\frac{1}{10000} \sum_{s=1}^{10000} (f^{(s)}(\widehat{\mathbf{d}}^{(m)}_{k}, z_k) - f(\widehat{\mathbf{d}}_k, z_k))^2 \right]^{\frac{1}{2}}.
\end{equation}
We employ $f(\widehat{\mathbf{d}}_k, z_k)$ here rather than $f_{opt,k}$ to understand how well the algorithms capture $f$ at the recommended dose combination even if the recommended dose combination is not optimal. This is important for later phase studies which may utilize estimates of $f$ obtained at the recommended dose combination for sample size and power calculations. In a similar manner, we present the average absolute deviation of the posterior mean point estimates $E[f_k \mid \mathcal{D}, \widehat{\mathbf{d}}_k, z_k]$ from the true value of $f(\widehat{\mathbf{d}}_k, z_k)$ at each iteration. Note that the standard algorithm ignores the strata and so yields only a single recommended dose $\widehat{\mathbf{d}}_{k} = \widehat{\mathbf{d}}$, posterior distribution $p(f_k \mid \mathcal{D}, \widehat{\mathbf{d}}_{k}, z_k) = p(f \mid \mathcal{D}, \widehat{\mathbf{d}})$, and posterior mean $E[f_k \mid \mathcal{D}, \widehat{\mathbf{d}}_{k}, z_k] = E[f \mid \mathcal{D}, \widehat{\mathbf{d}}]$ per iteration for $k=1,...,K$. See panels B-D of Figures \ref{fig:sim_scenario_1}-\ref{fig:sim_scenario_3} for these criteria by iteration.

Scenario 1 considers the case of no response heterogeneity across a single binary covariate $Z_1$ for small standardized effect sizes with monotonically increasing dose-efficacy surfaces.  Under this scenario, both algorithms converge to the locations of $\mathbf{d}_{opt,k}$, coming within one dosing unit of the optimal (Panel B of Figure \ref{fig:sim_scenario_1}). Estimation of $f$ is challenging for the small standardized effect size (i.e., higher level of noise), but by termination, the algorithms come within 0.4 units of the true $f$ at the recommended dose combination (panels C-D of Figure \ref{fig:sim_scenario_1}). The personalized algorithm is slightly less efficient than the standard algorithm, however, and takes longer to converge.  This is likely the result of the GP model used in the personalized algorithm needing to estimate the additional length-scale parameter for $Z_1$, $l_{1}$.

Scenario 2 considers response heterogeneity across $Z_1$ for large standardized treatment effect sizes with non-monotonic dose-efficacy surfaces. Under this scenario, the personalized algorithm converges to the locations of $\mathbf{d}_{opt,k}$ and estimates the true value of $f$ at the recommended dose combination well, whereas the standard algorithm does not (panels B-D of Figure \ref{fig:sim_scenario_2}). The standard algorithm typically explores the area near only a single $\mathbf{d}_{opt,k}$ or attempts to explore both optima (not shown). This results from the marginal objective function surface being bimodal, since it is a mixture distribution comprised of the equally weighted strata of $Z_1$, which are displayed in panel A of Figure \ref{fig:sim_scenario_2}. Note that even if the bimodality of the marginal surface is properly identified and explored, patients cannot be optimally treated without consideration of $Z_1$.  That is, without this additional covariate information, the standard algorithm cannot determine which mode should be used to treat patients with $Z_1=0$ versus $Z_1=1$.  This is possible using the personalized approach, however.

Scenario 3 considers heterogeneity across two binary covariates, $Z_1$ and $Z_2$, and includes zero, small, medium, and large standardized effect sizes with both monotone and non-monotone dose-efficacy surfaces.  Recall that the stratum $(z_1=0, z_2=0)$ corresponds to those patients who do not respond to the drug. Thus there are no optima in this stratum and the corresponding plot in panel B of Figure \ref{fig:sim_scenario_3} is empty. Panels C and D of Figure \ref{fig:sim_scenario_3} are not empty for this stratum, however, since $f=0$ everywhere, and so it is of interest to see how the algorithms estimate this value. Under this scenario, the personalized algorithm converges to the locations of $\mathbf{d}_{opt,k}$, coming within 1-1.5 dosing units of the optimal depending on the standardized effect size, whereas the standard algorithm does not (Panel B of Figure \ref{fig:sim_scenario_3}). Since the standard algorithm targets the global optimum, it performs best in stratum $(z_1=1, z_2=1)$ where the standardized effect is the largest. Accurate estimation of $f$ is challenging under this scenario (Panels C and D of Figure \ref{fig:sim_scenario_3}). The personalized algorithm shows evidence of convergence toward the true values of $f$ (i.e., RPSEL and absolute deviation of the posterior mean estimates decreasing to 0) whereas the standard algorithm does not. The standard algorithm yields only a single estimate of $f$ and so must split the difference among the different objective function values across the strata. This scenario suggests that as the number of strata grow, and thus also the likelihood for some degree of response heterogeneity to be present, the performance of the standard algorithm will be further degraded.

In summary, when heterogeneity exists across the strata, the personalized algorithm is superior in both identifying the locations of the $\mathbf{d}_{opt,k}$ and estimating $f$. When no heterogeneity exists, the standard algorithm is slightly more efficient. Additionally, the proposed methods have performed well for both monotonic and non-monotonic dose-efficacy surfaces, and have done so without utilizing strong prior information.

\section{Dose-Finding Design for an Intraocular Implant} \label{sec:rwa}

In this section we focus on the intraocular implant example.  The goal is to develop an intraocular implant with an optimal dose combination of two agents which reduce intraocular pressure (IOP), a laboratory confirmed measurement. The normal range of IOP is 12-21 mmHg, with 21-30 mmHg considered elevated IOP. Elevated IOP is a risk factor for ocular hypertension and glaucoma, and is strongly associated with increased vision loss \citep{leske2003factors}. The implant seeks to reduce elevated IOP by combining two agents, with doses $d_1$ and $d_2$, each of which has been in use individually in topical form for many years. The agents are well tolerated and we do not expect any drug-related adverse events. We are interested in obtaining the optimal dose combination of these two agents using reduction in IOP from baseline as a continuous efficacy measure. It is hypothesized that higher doses do not necessarily imply greater efficacy, which is expected to plateau or even decrease at higher levels of agent concentration. Additionally, we expect response heterogeneity to exist with respect to a particular characteristic of the lens of the eye, which we treat as a binary covariate $Z_1$, and are interested in a design which permits identification of potentially different optimal dose combinations according to this patient characteristic. 

We allow the dose-finding algorithm to explore the dose-combination region, assumed to be standardized, subject to the manufacturing precision constraint of $0.25$ standardized dosing units and deem it ethically permissible to proceed without safety-related dose escalation/de-escalation rules. It is hypothesized that the implant can reduce elevated IOP by 5 mmHg in individuals with $Z_1=0$, but may be even more effective in individuals with $Z_1=1$, leading to reductions as high as 10 mmHg. To assess the cost and size of a hypothetical trial, we are interested in comparing the standard and personalized dosing approaches under different stopping rule specifications for the scenario described above. The final design is then selected as the one which balances good performance while controlling expected cost.  Costs are measured in terms of enrolled participants and also the number of unique dose combinations, since there are engineering costs associated with production of novel doses. 

\begin{figure}
    \centering
    \includegraphics[width=\linewidth]{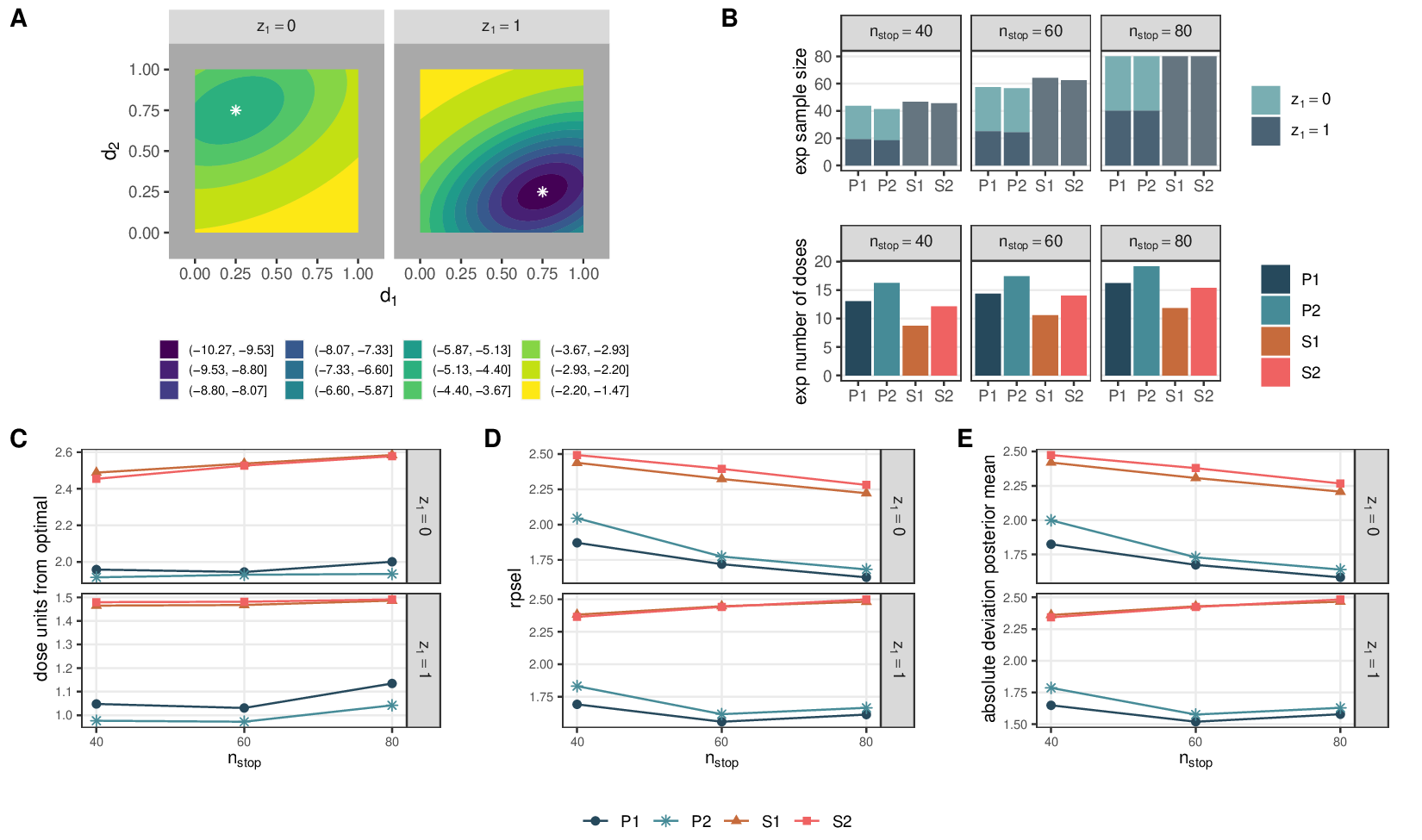}
    \caption{Intraocular Implant Scenario: 
    A) Objective function, white stars denote $\mathbf{d}_{opt,k}$, B) expected sample size and expected number of unique doses evaluated, C) expected dose units from the optimal dose combination as defined in (\ref{dose_unit_distance}) by iteration, D) average RPSEL as defined in (\ref{rpsel}) by iteration, and E) average absolute deviation of the posterior mean estimate by iteration. On the x-axis in (B), P stands for personalized and S for standard, followed by the number referring to high replication for a smaller number of doses (1) or low replication for a larger number of doses (2).}
    \label{fig:rwa_large_resp_het}
\end{figure}

The goal is to minimize the objective function. The data generating mechanism is $y = f(\mathbf{d},z) + \epsilon$ where $\epsilon \sim N(0, \sigma^2_y)$, with the specification of $f(\mathbf{d},z)$ included in the second panel of Table \ref{tbl:sim_study_scenarios} (row labeled ``Implant") and plotted in panel A of Figure \ref{fig:rwa_large_resp_het}.  We use the same indexing for $\mathbf{d}_{opt,k}$ and $f_{opt,k}$ as described previously. The value of $\sigma_y=5$ ensures medium standardized effect sizes of 1 and 2 for $Z_1=0$ and $Z_1=1$, respectively, across non-monotonic dose-efficacy surfaces. 

Two settings of the algorithms are compared for a maximum sample size of 80: one setting includes a higher number of replications at a smaller number of doses, and the other includes a smaller number of replications at a larger number of doses. We denote the standard/personalized algorithms under the first setting as $S1/P1$ and under the second setting as $S2/P2$. Under the first setting, $S1/P1$ are run under the same specifications described in the previous section, where $r=4$ and $r=2$ for the standard versus personalized algorithms, respectively. Under the second setting, $S2/P2$ are run with $r=2$ and $r=1$ for the standard versus personalized algorithms, respectively, for $c=10$ initial dose combinations selected via Sobol sequences. 

As the sponsor is concerned about cost and size of the trial, early stopping is permitted using the rule defined in (\ref{stopping_rule}). Early stopping is investigated by choosing values of $\delta$ as previously described such that there is a moderately high chance of stopping after roughly 40 or 60 total participants are enrolled in the trial (denoted by the values of $n_{stop}$ in Figure \ref{fig:rwa_large_resp_het}). These values are $\{0.00179, 0.000971\}$ for $S1$, $\{0.00670, 0.00345\}$ for $P1$, $\{0.00140, 0.000820\}$ for $S2$, and $\{0.00565, 0.00298\}$ for $P2$. Since we are considering a dual-agent dose combination, $J=2$ and we thus require the stopping criteria in (\ref{stopping_rule}) to be satisfied $J+1=3$ times before stopping early. 

For the personalized algorithm, we permit stratum specific early stopping.  Importantly, should exploration of one stratum stop early, we allocate the remaining budget to recruitment of participants in the other stratum. This assumes the sponsor can target recruitment specifically for this group.  This reallocation enables resources to be utilized in strata which are harder to optimize, and so may increase performance. We compare this to no early stopping ($i.e., n_{stop} = 80$), where the numbers of participants enrolled in each stratum are equal. When combined with the two settings for each algorithm, 12 unique designs are defined: $P1/P2/S1/S2$ each of which has three stopping rules, denoted by $n_{stop} = \{40,60,80\}$. All computing and inference is performed as previously described. The performance of the designs is compared using the previously defined criteria which are estimated via 1,000 Monte Carlo replicates.

The expected sample size and expected number of unique doses evaluated by each design for the scenario described above is included in panel B of Figure \ref{fig:rwa_large_resp_het}.  The standard and personalized designs have approximately equal expected samples sizes within each stopping rule, but the personalized designs are expected to evaluate more unique doses on average. The performance of the personalized designs with respect to expected dosing units from the $\mathbf{d}_{opt,k}$ (panel C of Figure \ref{fig:rwa_large_resp_het}) is comparable within each stopping rule, having differences of around 0.1 standardized dosing units, which are too small to be practically meaningful. 

There is some suggestion of a possible increase in the expected number of dosing units from $\mathbf{d}_{opt,k}$ under no early stopping (i.e., $n_{stop} = 80$) as compared to early stopping. To investigate this, an additional simulation was performed and compared designs with maximum samples sizes of 40 and 60 under no early stopping to those which permit early stopping at roughly these same number of participants.  The simulation suggests that the increase mentioned above results from the equal allocation of participants across strata under no early stopping. When early stopping is permitted, a larger proportion of evaluations is allocated to the stratum which is more difficult to optimize, and so provides an improved model fit at each iteration of the algorithm which improves the dose-finding overall. This difference in proportions can be observed in panel B of Figure \ref{fig:rwa_large_resp_het} by noting that higher proportions of the expected sample sizes under early stopping rules come from stratum $Z_1=0$, which has a smaller standardized effect size and is thus harder to optimize. Regardless, under the current scenario, the observed difference in expected dosing units from $\mathbf{d}_{opt,k}$ between the stopping rules $n_{stop}=60$ and $n_{stop}=80$ for the personalized designs is too small to be meaningful. However, future work should more fully investigate how equal versus unequal allocation of participants across the strata at each algorithm iteration impacts design performance.

Design $P1$ estimates $f$ the best, supporting findings in the literature that suggest higher degrees of replication can beneficial for estimation under noisy settings \citep{binois2018practical}. This difference is most apparent between designs $P1$ and $P2$ under stopping rule $n_{stop} = 40$ (panels D-E of Figure \ref{fig:rwa_large_resp_het}). The standard algorithms perform poorly across all performance metrics, recommending doses which are on average farther away from $\mathbf{d}_{opt,k}$, and poorly estimating the objective function $f$ at the recommended dose combinations (panels C-E of Figure \ref{fig:rwa_large_resp_het}). This poor performance is expected since response heterogeneity is present in the true data generating mechanism. If response heterogeneity were not present, we would expect the standard algorithms to be slightly more efficient than their personalized counterparts as was observed in the simulation study from the last section.

To suggest a final design to the sponsor, we use Figure \ref{fig:rwa_large_resp_het} as a visual aid. Since response heterogeneity is expected a priori, the poor performance of the standard designs under this scenario renders them inappropriate. Instead, we select the personalized $P1$ design with $n_{\text{STOP}}=60$.  For roughly the same expected sample size but for fewer unique dose evaluations, this design yields final dose suggestions which are as close to $\mathbf{d}_{opt,k}$ as design $P2$.  This design also offers a mild improvement in the estimation of $f$ as compared to $P1$ with $n_{stop} = 40$. Choosing $P1$ with ${n_{stop}=60}$ over $P1$ with ${n_{stop}=40}$ does come with additional cost, however: the design with $n_{stop}=40$ expects to evaluate 13 unique doses and enroll approximately 44 participants, whereas the design with $n_{stop}=60$ expects to evaluate about 15 unique dose combinations (a 15\% increase) and enroll approximately 58 participants (a 32\% increase). The sponsor would need to weigh the increased engineering and enrollment costs against the increase in performance.

\section{Discussion} \label{sec:discussion}

In this manuscript, we proposed the use of Bayesian optimization for early phase multi-agent dose-finding trials in a tolerated toxicity setting. We showed the benefit of taking a personalized approach for dual-agent trials when heterogeneity exists across a set of prespecified subgroups.  As expected, under no response heterogeneity the personalized approach is slightly less efficient. As noted in the introduction, parametric models may suffer from the curse of dimensionality when transitioning from standard to personalized dose-finding, as they require terms for all higher order dose-covariate interactions. By using the anisotropic squared exponential GP kernel in the Bayesian optimization methods proposed here, however, only a single additional parameter per covariate is required (the additional lengthscale parameter corresponding to that covariate). Thus, the proposed methods highlight the benefit and feasibility of adopting a personalized approach toward early phase multi-agent dose-finding trials for both monotonic and non-monotonic dose-response surfaces. 

The proposed approach is not without limitations.  First, the methods proposed in this work assumed no meaningful toxicity across the dose combination space. Extension to higher-grade toxicity settings through incorporation of dose escalation/de-escalation remains as future work. Second, the personalized approach was demonstrated by considering dual-agent dose combinations in predefined subgroups only. Extension to dose combinations with more than two agents is trivial, but extension to continuous covariates without categorization is not, and could be the subject of future investigations. Another direction for future development is to extend the proposed approach to binary and ordinal outcomes where the proposed response models may be defined over a latent continuous surface. 

In this manuscript, Bayesian optimization is utilized as a global optimizer. While other global optimization methods exist (e.g., genetic algorithms and simulated annealing), they require many function evaluations and are thus not appropriate for early phase dose-finding trials where evaluations are expensive \citep{bull_bayesopt, tracey2018upgrading}.  We employed the AEI acquisition function under a GP surrogate model. Performance of the algorithms under additional acquisition functions, surrogate models, and/or kernel functions should be investigated. The (stationary) separable anisotropic kernel used in this work assumes that all strata have the same correlation structure and that the covariance between points in different strata are changed by a multiplicative factor only, which may not be true in general. Indeed, under simulation scenario 3 which included two binary covariates, the objective function is zero everywhere for stratum $(z_1=0,z_2=0)$, so this assumption is not true in this case. The objective function values in this stratum are perfectly correlated, whereas those corresponding to dose combinations in other strata are not. Future work should investigate relaxing the assumption of stationarity by using kernels that are non-separable (e.g., including dose-covariate interaction terms in the kernel function (\ref{cov_squared_exp_kernel})), or even non-stationary, or deep GPs \citep{sauer2023active}. Finally, if the number of included covariates is large and there is reason to believe that only low level interactions between the drug combinations and covariates exist, different surrogate models could be employed, such as additive GPs \citep{duvenaud2011additive} or Bayesian additive regression trees \citep{chipman2010bart}. 

Finally, we adopted an empirical Bayes approach toward the GP hyperparameter estimation to decrease the computational burden of the simulations. Likelihood methods can yield poor results when the sample sizes are small, as in early phase dose-finding trials, and so full Bayesian inference may be preferred \citep{bull_bayesopt, wang2014theoretical}.  Unfortunately, the Markov chain Monte Carlo methods typically used to perform full Bayesian inference are prohibitively expensive for the algorithms proposed here, and so a sequential Monte Carlo approach may be a less computationally demanding alternative \citep{gramacy2011particle}. 

\section{Data Availability}
No new data were created or analyzed in this study. The R scripts used for the simulations and graphics can be found on a public GitHub repository at \url{https://github.com/jjwillard/bayesopt_pers_df}.

\section{Acknowledgements}
JW acknowledges the support of a doctoral training scholarship from the Fonds de recherche du Québec - Nature et technologies (FRQNT) and a research exchange internship with PharmaLex Belgium. SG acknowledges the support by a Discovery Grant from the Natural Sciences and Engineering Research Council of Canada (NSERC) and a Chercheurs-boursiers (Junior 1) award from the Fonds de recherche du Québec - Santé (FRQS). EEMM acknowledges support from a Discovery Grant from NSERC. EEMM is a Canada Research Chair (Tier 1) in Statistical Methods for Precision Medicine and acknowledges the support of a chercheur de mérite career award from the FRQS. This research was enabled in part by support provided by Calcul Québec and the Digital Research Alliance of Canada.

\bibliographystyle{apalike}
\bibliography{reference}

\end{document}